\documentclass[sigplan,10pt]{acmart}

\setcopyright{none}
\renewcommand\footnotetextcopyrightpermission[1]{}
\let\citestyle\relax
\expandafter\let\csname ver@cite.sty\endcsname\relax
\usepackage{amsmath,amsthm}
\usepackage{algorithm}
\usepackage{algpseudocode}
\usepackage{comment}
\usepackage[inline]{enumitem}
\usepackage{xcolor}
\usepackage{array}
\usepackage{url}
\usepackage{xurl}
\usepackage{xspace}
\usepackage{soul}
\soulregister\cite7
\usepackage{subcaption}
\usepackage{pifont} 

\newcommand{\one}{\ding{182}\xspace}
\newcommand{\two}{\ding{183}\xspace}
\newcommand{\three}{\ding{184}\xspace}

\definecolor{Gray}{gray}{0.85}
\newcolumntype{C}[1]{>{\centering\arraybackslash}p{#1}}
\newcolumntype{G}[1]{>{\centering\arraybackslash\columncolor{Gray}}p{#1}}
\newcolumntype{L}[1]{>{\raggedright\arraybackslash}p{#1}}
\newcolumntype{R}[1]{>{\raggedleft\arraybackslash}p{#1}}
\newcolumntype{D}[1]{>{\centering\arraybackslash} m{#1}}

\newcommand{\pagecombo}[2]{\texttt{G{#1}:H{#2}}\xspace}
\newcommand{\fourktwom}{\pagecombo{4K}{2M}}
\newcommand{\twomtwom}{\pagecombo{2M}{2M}}

\usepackage{xpatch}
\xpretocmd{\citeauthor}{\hypersetup{citecolor=black}}{}{}
\newcommand{\ncite}[1]{{\citeauthor{#1}}~\cite{#1}}

\newcommand{\sepblock}{
\smallskip
\noindent
}

\newcommand{\nearmemory}{{near memory}\xspace}
\newcommand{\farmemory}{{far memory}\xspace}
\newcommand{\quotes}[1]{``#1''}

\newcommand{\mname}{GPAC\xspace}
\newcommand{\filter}{{\it Scattered Page Filter}\xspace}
\newcommand{\consolidator}{{\it Page Consolidator}\xspace}
\newcommand{\limit}{{\tt CL}\xspace}

\newcommand{\skewedhotpage}{skewed hot huge page\xspace}
\newcommand{\skewedhotregion}{skewed hot huge region\xspace}
\newcommand{\cxl}{{CXL}\xspace}
\newcommand{\ipt}{{IPT}\xspace}
\newcommand{\memtierd}{Memtierd\xspace}
\newcommand{\baseguest}{{\sf G:4K}\xspace}
\newcommand{\hugeguest}{{\sf G:2M}\xspace}
\newcommand{\hostOne}{\textsf{Host1}\xspace}
\newcommand{\hostTwo}{\textsf{Host2}\xspace}
\newcommand{\guestOne}{\textsf{Guest1}\xspace}
\newcommand{\guestTwo}{\textsf{Guest2}\xspace}

\newcommand{\newtext}[1]{{\leavevmode{#1}}}

\setlist[enumerate]{itemsep=0pt, leftmargin=11pt, topsep=2pt}
\begin{document}

\title{Efficient Memory Tiering in a Virtual Machine}

\author{Chandra Prakash}
\affiliation{%
  \institution{Intel}
  \country{India}
}

\author{Sandeep Kumar}
\affiliation{%
  \institution{Intel}
  \country{India}
}

\author{Aravinda Prasad}
\affiliation{%
  \institution{Intel}
  \country{India}
}

\author{Sreenivas Subramoney}
\affiliation{%
  \institution{Intel}
  \country{India}
}


\pagestyle{plain}

\begin{abstract}
Memory tiering is a well established technique to tackle the increasing server memory total cost of ownership (TCO) and growing memory demands of data center workloads. A tiering solution dynamically places frequently accessed data (hot) to a small, fast, expensive memory tier known as near memory and  infrequently accessed (cold) data to a low-cost, slow, capacity memory tier known as far memory. Precise identification of hot and cold data is the key to efficient tiering.
However, in a virtualized setup where the host uses huge pages,  host-level tiering can be inefficient when (i) hot base pages are \textit{scattered} across guest physical address space, mapping to many host huge pages, or (ii) accesses within a guest huge page are \textit{skewed} to a few base pages. In both cases, the host sees entire corresponding mapped huge pages as hot, and places them in near memory, wasting the precious resource.

We propose GPAC, a guest-side technique that exploits the two-level address translation to consolidate scattered and skewed hot base pages into a compact set of guest physical address ranges. This transforms many skewed hot huge pages into a few densely hot huge pages at the host level, enabling existing host-based tiering solutions to use near memory effectively. GPAC is host-agnostic, requires no hardware or hypervisor modifications, and is compatible with any host-based tiering solution.
Our evaluation with DRAM as near memory and CXL as far memory shows 50--70\% reduction in near memory consumption at similar performance levels with standalone real-world benchmarks and state-of-the-art host-based tiering. At scale, with multiple guests, GPAC improves performance by 4--11\% with similar memory TCO.

\end{abstract}

\maketitle


\section{Introduction}
\label{sec:intro}

Memory tiering~\cite{gswap,tmo,autotiering,tpp,hemem,memtis,mtm,tmts} on a heterogeneous memory system has emerged as a practical way to address the increasing cost of memory in data centers, which now accounts for 33--90\% of the total cost of ownership (TCO)~\cite{vmware_tco,tmo, src-study} and is projected to grow further with demands of modern data-hungry applications~\cite{llm3,llm2,llm}.
Memory tiering solutions aim to strike a balance between performance and TCO savings. A tiering solution periodically identifies frequently accessed (hot) data pages and places them in a small, fast, and expensive ``near memory'' tier (HBM, DRAM). It also moves infrequently accessed (cold) data pages to a large, slow, and cost-effective ``far memory'' tier (CXL, NVMM).

\begin{figure}
    \centering
    \includegraphics[width=.9\linewidth]{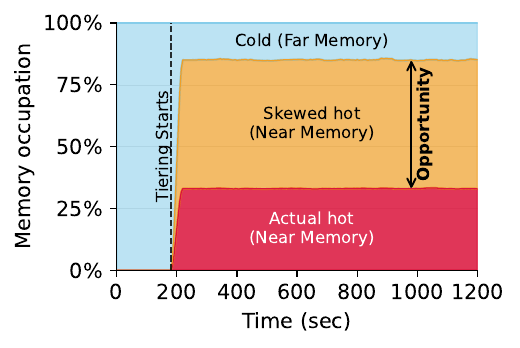}
    \caption{In a virtualized setup with host and guest using huge pages, $\approx$52\% of Redis's total memory footprint (running inside the guest) is incorrectly identified as hot by an host-based tiering technique using ideal page tracking, and is placed in the \nearmemory (fast memory). 
    }
    \label{fig:teaser}
\end{figure}

In modern data centers operating in a virtualized setup, memory tiering is employed at the host level.
Cloud Service Providers (CSPs) enable huge pages at the host due to their performance benefits~\cite{efficient_ab,weiwei}, while guests are free to use either huge pages or base pages. The guest physical address (GPA) has a one-to-one mapping with the host virtual address (HVA), and host-level tiering tracks hotness at HVA granularity, which corresponds to a huge page at the host.

However, host-based tiering can be rendered ineffective when \one in a guest using huge pages, accesses within a guest huge page are \textit{skewed} to a few base pages or \two in a guest using base pages, hot base pages are \textit{scattered} in the guest physical address space. Since guest pages are mapped to host huge pages, any access within the huge page boundary, either skewed or scattered, will make the corresponding huge page at host appear hot~\cite{memtis, hugescope}, resulting in several {\skewedhotpage}s at the host level. 
As memory tiering techniques are based on page hotness, skewed hot huge pages are placed in near memory, leading to under-utilization of costly near memory resources~\cite{vtmm,hugescope} and reducing opportunities to place other guests' hot huge pages in near memory.
As shown in Figure~\ref{fig:teaser}, for Redis workload running inside a VM and the host always using huge pages, $\approx 52\%$ of the total memory footprint is incorrectly identified as hot by a host-based tiering technique using ideal page tracking. All the identified hot pages are moved to \nearmemory. 

In a non-virtualized setup, \ncite{memtis} addresses this problem by using Intel PEBS~\cite{pebs} to track memory access at byte level and splits \skewedhotpage into base 4\,KB pages.
However, in a virtualized setup, hardware-assisted byte-level access tracking is not available for guests~\cite{pebslimit_avoid, hugescope}, and host-level tiering solutions must use page table-based memory access tracking, which only tracks at page granularity. The state-of-the-art, HugeScope~\cite{hugescope}, proposes identifying \skewedhotpage by splitting huge pages. Although this solves the issue of \skewedhotpage, it loses the benefit of huge pages at the host level. Moreover, HugeScope requires extensive modifications at the hypervisor level for efficient huge page splitting and coalescing.

We propose \textit{{G}uest {P}hysical {A}ddress {C}onsolidation} or {\mname},  a novel, configurable guest-side technique that enhances memory tiering at the host level by significantly reducing the number of {\skewedhotpage}s at the host while maintaining huge pages at the host level. GPAC is agnostic to the host-level tiering solution and memory tiers, as well as the telemetry technique inside the guest. It requires no support from the host and works without any hardware or hypervisor modifications. GPAC identifies {\skewedhotpage}s inside the guest and only demotes them within the guest if the guest is using huge pages; huge pages at the host are never demoted, allowing both the host and the guest to continue benefiting from huge pages.

\mname exploits the two-level address translation and consolidates hot base pages from multiple {\skewedhotpage}s into a few densely hot huge pages. This enables memory tiering techniques in the host to effectively place ``actual" hot huge pages in near memory, reducing  per-VM near memory consumption with minimal performance degradation.  GPAC only modifies the GVA to GPA mapping; the HVA to HPA mapping remains untouched, and the host continues to get benefits of huge pages.
To control the aggressiveness of consolidation and limit its impact on performance, GPAC introduces a tunable parameter called \textit{Consolidation Limit} or \limit that determines whether a hot huge page inside the guest is classified as skewed.  A higher \limit favors near memory savings at the cost of performance, while a lower \limit prioritizes performance over memory savings.



\noindent
Our main contributions are as follows:
\begin{enumerate}[leftmargin=3mm]
\item We propose \mname, a host-agnostic, guest-side consolidation technique that reduces the number of {\skewedhotpage}s at the host without splitting huge pages at host level. \mname requires no hardware or hypervisor modifications.
\item \mname introduces a tunable consolidation limit (\limit) that allows users to control the trade-off between near memory savings and performance.
\item Our evaluation with state-of-the-art host-level tiering solutions shows that \mname reduces near memory consumption by 50--70\% for a single VM with minimal performance overhead, and improves throughput by 4--11\% at scale with multiple VMs at similar memory TCO.
\end{enumerate}

\section{Background and Related Work}
\label{sec:bg_and_related_work}

\subsection{Two Dimensional Address Translation} 
\label{sec:bg_address_translation}
In a bare metal setting, translating  a virtual address (VA) to a physical address (PA) requires a single-level page table walk on a TLB miss. However, in a virtualized environment, a 2-dimensional (2D) walk or nested walk is required to convert the \textit{guest virtual address} (GVA) to the \textit{host physical address} (HPA).
The first level walks the \textit{guest page table} or GPT to translate GVA $\xrightarrow{GPT}$ GPA, while the second level walks the \textit{host page table} or HPT to translate HVA $\xrightarrow{HPT}$ HPA.  In virtualized environments such as QEMU/KVM~\cite{kvm}, the GPA has a one-to-one linear mapping with the \textit{host virtual address} (HVA). 
To reduce the translation overhead, hardware-assisted virtualization solutions like Extended Page Table (EPT)~\cite{ept} or Rapid Virtualization Indexing (RVI)~\cite{rvi} were developed to reduce address translation overhead. 
However, in the worst case, GVA to HPA translation requires 24 memory accesses with a 4-level page table and 35 in 5-level page table~\cite{page_walk_ept,translation_pass_through_atc23_35memaccess_5lvl}.

\sepblock
\textbf{Translation with huge pages:} 
A huge page is a contiguous group of base pages, virtually and physically. The supported huge page sizes depend on CPU and operating system support ( for example, huge pages of sizes 2\,MB and 1\,GB on Intel x86~\cite{huge_benefits}). In a virtualized setting, huge  pages increase TLB reach and reduce TLB miss latency by eliminating the last level of the page table walk.

\subsection{Telemetry --- Access Profiling}
\label{sec:bg_telemetry}
Generating an accurate access profile (page addresses and corresponding access counts) is the key to effective memory tiering~\cite{telescope}. State-of-the-art telemetry techniques either rely on page table entries~\cite{idle-page-tracking, damon, autonuma} or hardware counters~\cite{pebs} to build the access profile.

\sepblock
\textbf{Access Bit tracking:} These methods~\cite{damon,telescope,memtierd} rely on the manipulation and monitoring of \texttt{ACCESSED} ($A$) bit in the page table, which is automatically set by the hardware upon a page table walk~\cite{telescope}. A software daemon periodically clears this bit and checks it again after a certain time window. If the bit is set, that indicates that the page was used; otherwise, not. 

\sepblock
\textbf{NUMA Hints:} Another set of methods~\cite{autonuma,tpp} rely on the \texttt{PROT\_NONE} or $P$ bit to capture the access profile. A daemon sets the $P$ bit temporarily marking pages as inaccessible. A subsequent access to those pages triggers a minor fault. The access is recorded while serving the minor fault and later used to identify hot data pages. 

\sepblock
\textbf{Hardware counters:} In hardware-based telemetry, the hardware collects statistics about the memory access pattern based on certain events such as \texttt{LOAD} and \texttt{STORE} instructions. It collects the corresponding virtual memory addresses and reports them to the OS or the software. {Intel PEBS}~\cite{pebs} is a hardware-based mechanism to collect the memory access pattern of an application and has been used by different memory tiering solutions~\cite{hemem,memtis,autotiering}.

\subsection{Memory Tiering}
\label{sec:bg_memory_tiering}
Memory tiering is a widely adopted solution to address the growing cost of memory in data centers. CSPs employ a heterogeneous memory system with a fast, costly, and small-capacity memory tier such as DRAM or HBM, and a slow, cost-effective, and large-capacity memory tier such as CXL-attached memory~\cite{cxl, cxl-micron} and NVMM~\cite{nvmm}, henceforth referred to as \nearmemory and \farmemory, respectively. 
A tiering solution dynamically places the data across these different memory tiers based on the access pattern of the application such that most of the LLC misses are served from a \nearmemory while ensuring that cold data is placed on \farmemory. 
There is a plethora of memory tiering solutions proposed by academia and industry~\cite{gswap,tmo,autotiering,tpp,hemem,memtis,mtm,tmts}. 

These tiering solutions differ based on three parameters: \one the telemetry technique used to collect data page access information (\S\ref{sec:bg_telemetry}), \two the choice of memory tiers, such as compressed memory~\cite{tmo}, NVMM~\cite{autotiering,hemem,memtis,mtm,tmts}, or CXL-attached memories~\cite{autotiering,tpp,memtis}, and \three the policy used to promote or demote pages across tiers, which may rely on access counts~\cite{hemem,gswap}, fault counts~\cite{autotiering,autonuma,tpp}, or a combination of events~\cite{tmts}.

\sepblock
{\bf Memory tiering in virtualized environment:} 
In a virtualized setup, a memory tiering solution \newtext{can either be employed at the host with guests unware of the underlying memory tiers~\cite{heteroos,memstrata,vmware_host_based_tiering}, or delegrated to the guests~\cite{demeter}.
The tiering solution uses \texttt{ACCESSED} bits or \texttt{PROT\_NONE} bits (\S\ref{sec:bg_telemetry}) at HVA granularity, or EPT-friendly Intel PEBS to capure accesses within the guest~\cite{demeter} to build an access profile. The generated profile is then used to migrate pages across different memory tiers. }
\section{Motivation}
\label{sec:motivation}

\subsection{Access Skewness in a Huge page}
\label{sec:motiv_skewness_huge_page}

\begin{figure}[t]
    \centering
    \includegraphics[width=0.99\linewidth]{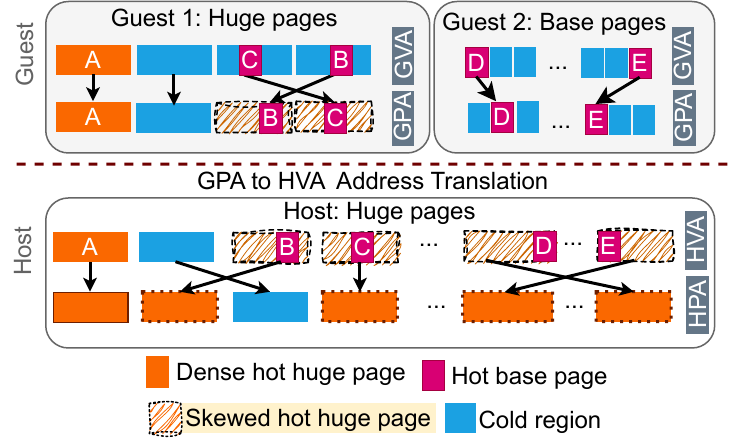}
    \caption{Address translation in a virtualized environment and the issue of {\it Skewed} hot huge pages. Host has four skewed hot huge pages: two from \guestOne and two from \guestTwo.}
    \label{fig:virt_huge_motivation}
\end{figure}

Huge pages alleviate pressure on the TLB and page caching structures, and are also recommended in virtualized environments (both guest and host) to improve application performance, as they reduce TLB pressure and Extended Page Table (EPT) walks on TLB misses~\cite{page_walk_ept, gemini_vmm}.
However, they introduce challenges for memory tiering in the presence of \skewedhotpage, where a few hot base pages can make the entire huge page hot~\cite{memtis}.
A hot huge page is considered {\it skewed} if the number of hot base pages is below a defined {\it skewness threshold}.  In a virtualized environment, CSPs typically enable huge pages at the host due to the aforementioned benefits of huge pages. Guests can either use base pages or huge pages 
which are backed by huge pages at the host.

As shown in  Figure~\ref{fig:virt_huge_motivation}, a \skewedhotpage in the host context can be due to either \one a \skewedhotpage inside the guest (mapped to a host huge page) or \two scattered hot base pages in the guest (mapped to different host huge pages). Figure~\ref{fig:virt_huge_motivation} explains skewness in detail with an example of two guests, \guestOne, which uses huge pages, and \guestTwo, which uses only base pages.
\guestOne frequently accesses one guest huge page A and two base pages within guest huge pages B and C, resulting in a total of three hot huge pages in both guest and host due to one-to-one GPA to HVA mapping. Out of the three hot huge pages, A is an actual hot huge page, whereas B and C are \skewedhotpage{s}.
\guestTwo frequently accesses two base pages (D and E) mapped to different host huge pages leading to two \skewedhotpage{s} at the host level. A  host-based memory tiering solution will place all five identified hot huge pages in the \nearmemory, even though only one (A) is actually hot and remaining four are \skewedhotpage, wasting precious resources.

\subsubsection{Skewness and Impact on tiering}
\label{sec:skewness}

\begin{figure}[t]
      \centering
      \includegraphics[width=0.45\textwidth]{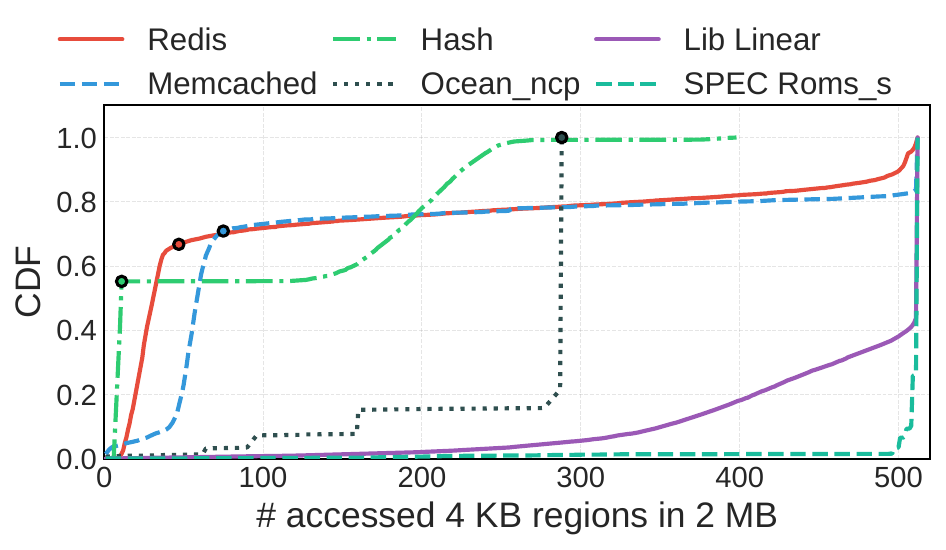}
      \caption{CDF of the number of accessed 4\,KB base pages within a 2\,MB huge page for a set of  real-world workloads. The dots represent the knee points and are discussed in Section \S\ref{sec:cl_knee} to be used by \mname.
      }
      \label{fig:scatter_cdf}
\end{figure}

We measure the amount of skewness present in a set of real world applications. Workload details and setup are provided in Table~\ref{tab:benchmarks} and Section~\ref{sec:exp_setup}, respectively.
Figure~\ref{fig:scatter_cdf} shows the CDF of the number of 4\,KB base pages accessed within each 2\,MB huge page across different workloads. For a given value on the x-axis, y-axis indicates the fraction of huge pages that have that many or fewer base pages accessed. In Memcached~\cite{memcached} (with Memtier benchmark~\cite{memtier_bench}), for $\approx85\%$ of 2\,MB huge pages,
fewer than 100 of their 512 4\,KB base pages are accessed. Hence, due to this {\it skewness}, 85\% of the total 2\,MB huge pages are marked hot even though less than 400\,KB of data within those individual huge pages is accessed. In contrast, workloads such as Liblinear~\cite{liblinear} and SPEC {\tt 654.Roms}~\cite{spec_roms} have most of their huge pages densely hot.

\sepblock
\textbf{Impact on tiering:} Host-based memory tiering techniques identify hot data pages of all the guest instances and place them in the fast memory tier. As these solutions cannot differentiate between a dense hot huge page and a \skewedhotpage, they end up placing \skewedhotpage{s} in the limited-capacity and costly \nearmemory tier. Our evaluation using Redis~\cite{redis} shows that $\approx52\%$ of host huge pages placed in \nearmemory were skewed (Figure~\ref{fig:teaser}).

\subsection{Limitations of State-of-the-art}
\label{sec:motiv_sota_limitations}
Prior work, HugeScope~\cite{hugescope}, proposes changes to hypervisor to detect \skewedhotpage by splitting huge pages. 
This splitting solves the issue of \skewedhotpage as the host-level tiering solution can now track the access profile at base page granularity.
%
All subsequent tiering operations such as migration from/to \nearmemory are performed at the base page granularity. The solution comes at a cost of increased TLB-related activities at the host level. 
{Moreover, HugeScope is built on top of vTMM~\cite{vtmm} for memory tiering at host which needs additional guest support.
Note that only splitting the guest huge page while retaining the corresponding huge page in the host only shifts the type of skewness from a skewed hot huge page in the guest to scattered hot base pages in the guest.
}

\newtext{Demeter\cite{demeter} enables EPT-friendly Intel PEBS-based tracking inside a guest and propose a design where the complete memory tiering is delgated to the guest -- accesst tracking, tiering policy, and page migrations.
However, the authors do note that, ``Demeter does not directly address intra-hugepage access skewness". Demeter maintains a minimum split granularity of 2MB page size to improve TLB efficiency and avoid management overhead.}

\mname, a novel approach, addresses the problem of skewness in the guest without splitting the huge pages at the host level and is compatible with any memory tiering solution at the host. \mname outperforms HugeScope by $\approx$4.2\% in throughput at scale without any hypervisor modifications (\S~\ref{sec:perf_vtmm_hugescope}).

\section{\mname Design and Implementation}
\label{sec:design}
    

In this section, we discuss the {\mname}'s scope, design goals and the key idea. We also discuss the design components and their implementation. \mname is applicable for any sized huge page. However, for ease of discussion we discuss assuming a  2\,MB page composed of 512 4\,KB base pages. 

\sepblock
\textbf{Huge page region:} The design of \mname is independent of whether a guest enables huge pages or not. We use the term \textit{huge page region} to keep the discussion applicable to both the scenarios. Inside a guest, if huge pages are enabled, a huge page region is the same as a guest huge page. Otherwise, it is a contiguous, GPA-aligned region of huge-page size with no guest-side TLB benefits. A huge page region is always backed by a same-size huge page at the host.

\subsection{Design goals}
\label{sec:scope_goal}
We design \mname to achieve the following goals:
\begin{enumerate}
    \item {\bf Host-agnostic:} Develop a solution that does not need any modification in the host or hypervisor\label{goal_host_agnostic}.
    \item {\bf Tiering technique agnostic:} Independent of any memory tiering solution at the host without requiring modifications\label{goal_tiering}. 
    \item {\bf Memory tier agnostic:} Independent of memory tiers, such as DRAM, Intel Optane, CXL-attached memory, High Bandwidth Memory (HBM).
    \item {\bf Telemetry agnostic:} Any supported telemetry technique (mentioned in \S\ref{sec:bg_telemetry}) can be used to identify \skewedhotpage in the guest.\label{goal_telemetry}.
    \item {\bf Leverage huge pages:} No splitting of huge pages at the host to get huge page benefits\label{goal_huge}.
\end{enumerate}

\subsection{{Scope}}
\label{sec:design_scope}
\mname aims to mitigate {\skewedhotpage}s at the host-level while operating completely in the guest. The input to \mname is a set of telemetry data (access counts) at base page granularity inside the guest. \mname is responsible for identifying {\skewedhotpage}s based on the telemetry data and mitigating the issue. \mname is \textit{not} responsible for generating the telemetry data. Any existing supported state-of-the-art technique can be used with \mname~\cite{vtmm, idle-page-tracking, tpp}. We assume that hosts always use huge pages to benefit from the higher TLB reach~\cite{google_cloud_huge_page,microsoft_cloud_huge_page,ingens} and are free to use any supported memory tiering technique.

\sepblock
{\bf GPAC placement in Cloud:}
In a typical setup, the guest is not aware of the available memory tiers and the memory tiering at the host~\cite{vmware_host_based_tiering}. However, since \mname is a guest-side technique, the cost of GPAC in terms of CPU cycles is paid by the guest. A natural question arises: what incentive does a guest have to bear this cost? We envision a cooperative model where CSPs offer GPAC-enabled VMs at a lower cost. GPAC enables better utilization of \nearmemory at the host-level. This allows CSPs to host additional VMs at similar performance while also reduces rental costs, creating a ``win-win'' for both guest and host.

\subsection{Key idea - Consolidation} 
\label{sec:design_key_idea}

A \skewedhotpage occurs at the host level when only a portion of a guest-mapped host huge-page is actively accessed. 
\mname exploits the additional level of address translation inside a guest (GVA $\xrightarrow{GPT}$ GPA) to \quotes{\textit{consolidate}} scattered hot regions at the granularity of 4\,KB base pages in the guest into a contiguous huge page size region inside the guest. Consolidation consists of \one data copy of scattered hot base pages and \two updating the $GVA\xrightarrow{GPT}GPA$ mapping accordingly. 
By relocating the hot data out of the huge page region, the corresponding host huge page appears as cold. Thus, reducing the number of {\skewedhotpage}s at the host. This operates entirely within the guest and requires no modification at the host (hypervisor) level. 

\begin{figure}[t]
      \centering
      \includegraphics[width=\linewidth]{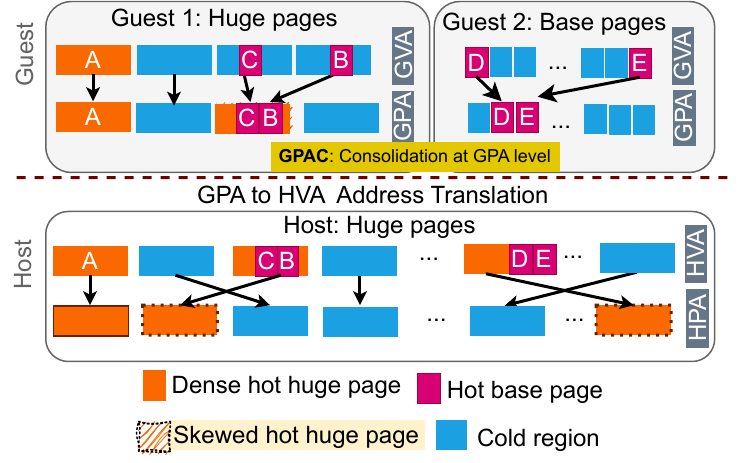}
      \caption{Hot base page consolidation inside guests to reduce the number of {\it skewed} hot huge pages at the host. All {\it skewed } hot huge pages are converted into dense hot huge pages.}
      \label{fig:virt_huge_solution}
\end{figure}

\noindent
\textbf{Example:} Using the same example from earlier (\S\ref{sec:motiv_skewness_huge_page}), as shown in  Figure~\ref{fig:virt_huge_solution}, consolidation moves the content of hot 4\,KB base pages $B$ and $C$ together in a contiguous huge-page sized region inside the guest (backed by an actual huge page at host) and modifies $GVA \xrightarrow{GPT} GPA$ mappings inside {\tt Guest-1}.
Similarly, pages $D$ and $E$ inside {\tt Guest-2} are also ``consolidated" to a single huge page region.
As shown in Figure~\ref{fig:virt_huge_solution}, after consolidation, all {\skewedhotpage}s at the host level are eliminated, resulting in three densely hot huge pages instead of  five hot huge pages earlier, thus, reducing the amount of data in \nearmemory.

\subsection{\mname components}
\label{sec:design_overview}

\begin{figure}[t]
      \centering
      \includegraphics[scale=0.8]{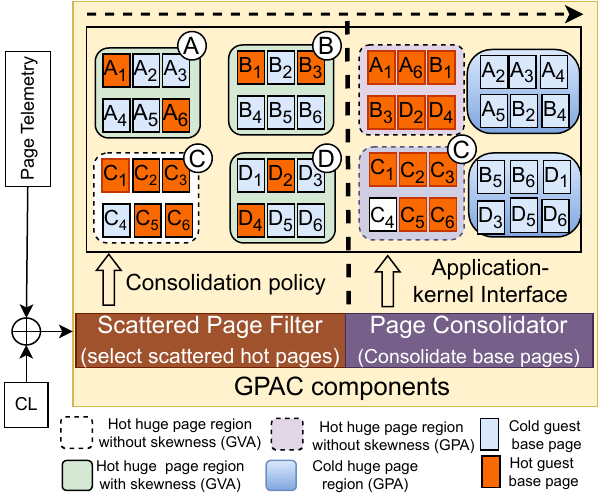}
      \caption{Design components of \mname for memory consolidation inside guest. {\it Scattered Page Filter} selects huge page region A, B, and D skewed based on CL value as 3. {\it Scattered Page Filter} pass the addresses $A_1$, $A_6$, $B_1$, $B_3$, $D_2$, and $D_4$ to {\it Page Consolidator} for consolidation at GPA level.
     }
      \label{fig:high_level}
\end{figure}

At a high level, GPAC's approach consists of the following steps. 
The \one first step is to collect telemetry data, i.e., access profile of data within guest at base page granularity. The \two second step is to apply a ``policy" and identify {\skewedhotpage}s inside the guest. The \three third step is to address the {\skewedhotpage}s inside the guest.  \mname operates through guest user and kernel space and consists of two major components inside the guest: \filter and  \consolidator as shown in Figure~\ref{fig:high_level}. 

\subsubsection{Scattered Page Filter}
\label{sec:scattered_memory_filter}
\filter operates in the guest's address space and is responsible for identifying {\skewedhotpage}s inside the guest based on an access profile at base page granularity as an input and the value of a user-tunable knob we call \textit{consolidation limit} or \limit. Consolidation reduces the number of {\skewedhotpage}s at the host but incurs performance overhead due to data copy and page table updates inside the guest. \mname uses \limit to balance memory savings and performance overhead due to consolidation.

\sepblock
\textbf{Consolidation candidate :}
GPAC profiles the application for a given time window $T$ and collects the access profile using the preferred telemetry method.  Based on the profile, GPAC counts the number of hot 4\,KB base pages $h_i$ for huge-page region $i$. All huge page regions with  $ h_i < \limit$ are classified as {\skewedhotpage}s and are candidates for consolidation. 
The candidate set is:
\begin{equation}    
\mathcal{C}(CL)=\{\, i \mid h_i < CL \,\}.
\end{equation}

\sepblock
\textbf{Memory savings and Work Done:}
For a highly skewed hot huge region (small $h_i$) consolidation offers large memory savings ($512 - h_i$ cold pages) at low cost ($h_i$ consolidations). On the other hand, for a low skewed hot huge region (high $h_i$) the memory savings are low and the consolidation cost is high.
Given a candidate set for a particular \limit, the total memory savings and total work done are defined as:
\begin{equation}
S(CL) = \sum_{i \in \mathcal{C}(CL)} (N - h_i),
\qquad
W(CL) = \sum_{i \in \mathcal{C}(CL)} h_i,
\end{equation}
Here, $N$ is the number of base pages inside a huge page ($N$=512 for a  2\,MB huge page).
We define the \emph{effectiveness} of consolidation as:
\begin{equation}
    E(CL) = S(CL) - \lambda\,W(CL),
\end{equation}
where $\lambda>0$ is the per-page migration cost and is system- and workload- dependent. $E(CL)$ captures the net efficiency of consolidation for a given \limit. A low value of $\lambda$ indicates the migration cost is low and favors an aggressive consolidation. A high value of $\lambda$ indicates high migration cost and favors a conservative consolidation. \newtext{Note that} estimating $\lambda$ directly is non-trivial, as it depends on the memory tier, memory pressure, bandwidth, and workload access pattern.

\subsubsection{Default CL Values}
\label{sec:cl_knee}
 To approximate a practical default \limit, we apply the ``Kneedle'' algorithm~\cite{kneedle} to the per-workload CDF of $h_i$ across its huge page regions (see Figure~\ref{fig:scatter_cdf}). The detected ``knee'' point captures the transition from highly skewed to low-skewed regions, beyond which savings diminish relative to the work required. We use this knee as the default \limit, capped at 60\% of $N$ ($N=512$ for 2\,MB) based on our empirical observation, as consolidation beyond this point yields diminishing returns. 

\sepblock
\textbf{CL tunability:} 
CL is fully tunable in the range $[1,511]$ assuming a 2\,MB huge page. A user can decrease CL to limit consolidation overhead at the expense of smaller memory savings or increase it beyond the default values to increase the memory savings at higher consolidation cost. We evaluate the impact of different values of CL on memory savings and performance in \S\ref{sec:cl_impact}.

\subsubsection{Page Consolidator}
\label{sec:design_consolidator}
Once the consolidation candidate set of {\skewedhotpage}s is identified, the \consolidator, running in the guest's kernel-space, demotes each candidate to base pages and consolidates the hot base pages into densely hot huge page regions. The set of {\skewedhotpage}s can serve as target regions where hot and cold base pages are swapped in place, without fresh page allocation as shown in Figure~\ref{fig:high_level}. Each swap requires three copy operations but opens optimization opportunities to reduce the total number of swaps.

Alternatively, the consolidator processes hot base pages in batches of up to 512. For each batch, it allocates a contiguous huge-page sized  memory region, copies the content of each hot base page into the new region, and updates the corresponding GVA to GPA mappings. Since the GVA remains unchanged from the application's perspective, the consolidation process is transparent to workloads. The newly allocated regions are not guest huge pages; they are huge-page sized regions at base page granularity. However, they are aligned and backed by actual huge pages at the host (\S\ref{sec:design_key_idea}). To ensure that the region allocation succeeds under memory pressure, a small pool of huge-page-sized regions can be pre-reserved. 



\subsection{Implementation}
\label{sec:implementation}

\newcommand{\code}[1]{\texttt{#1}}

\begin{algorithm}[t]
\footnotesize
\caption{Page Consolidation: {\tt consolidate\_pages()} }
\label{algo:page_consol}
\begin{algorithmic}[1]
\Require $L$: List of hot base pages from identified skewed hot huge regions.
\While{$L$ is not empty}
    \State $B \gets$ remove up to 512 pages from $L$
    \Statex \hspace{\algorithmicindent} \textbf{Allocate target region:}
    \State $HR_{ptr} \gets \texttt{page\_alloc(2\,MB)}$
    \If{$HR_{ptr}$ is {\tt NULL}}
        \State \Return {\tt -ENOMEM}
    \EndIf
    \Statex \hspace{\algorithmicindent} \textbf{Migrate each page in batch:}
    \For{each {\tt old\_page} in $B$ with index $i$}
        \State {\tt new\_page} $\gets HR_{ptr} + i \times 4\,\text{KB}$
        \State {\tt addr} $\gets$ virtual address of {\tt old\_page}
        \State \code{lock\_page(old\_page), lock\_page(new\_page)}
        \State {\tt ptep} $\gets$ \code{get\_pte(addr)}
        \State \newtext{\code{unmap(old\_page)}} \Comment{Unmap the page}
        \State \newtext{\code{flush\_tlb\_mm\_range(mm, addr, addr + 4\,KB)}} 
        \State \code{memcpy(new\_page, old\_page)} 
        \State {\tt new\_pte} $\gets$ \texttt{mk\_pte(new\_page, ...)}
        \State \code{set\_pte\_at(mm, addr, ptep, new\_pte)} \Comment{Remap GVA}
        \State \code{unlock\_page(new\_page), unlock\_page(old\_page)}
        \State \code{free(old\_page)} \Comment{Release old page}
    \EndFor
\EndWhile
\end{algorithmic}
\end{algorithm}

\subsubsection{Scattered Page Filter}
\label{sec:impl_filter}
\filter requires an access profile at base page granularity to identify {\skewedhotpage}s. 
\newtext{\mname is independent of the telemetry technique and can seamlessly work with either page-table-based tracking or recently introduced Intel PEBS-based tracking within the guest~\cite{demeter}. 
However, the Intel-PEBS support is not yet in the mainline kernel. We tried running the artifact from Demeter~\cite{demeter} on our Emerald Rapid Server. However, due to technical challenges, it did not work. Hence, we implement \mname using page table-based tracking. Here, \mname uses the Idle Page Tracking tool~\cite{idle-page-tracking} (or IPT) as its telemetry technique inside the guest to build an access profile in a given time window. Due to the limitation of the tracking tool, we configure the guest to use base 4\,KB pages ( as IPT will not be able to give 4\,KB access information, crucial to identify \skewedhotpage, if the guest continues to use huge pages).
The host continues to use huge pages.



}

\subsubsection{Page Consolidator}
\label{sec:impl_consolidator}
\consolidator exposes an application-kernel interface to get necessary information from the guest user space, such as process {\tt pid} and the candidate set of {\skewedhotpage}s to be consolidated (see Algorithm~\ref{algo:page_consol}).
The interface is implemented using a custom system call \quotes{{\tt consolidate\_pages()}} which takes the candidate set and their corresponding hot base pages. 
It then consolidates the hot base pages by using the allocation mechanism discussed in Section~\ref{sec:design_consolidator}.


\section{Evaluation}
\label{sec:eval}

\subsection{Experimental setup}
\label{sec:exp_setup}

\subsubsection{Host configuration}
Table~\ref{tab:hosts} summarizes our testbed configurations. We use two hosts, \hostOne and \hostTwo, each equipped with different memory tiers and host-based memory tiering to evaluate the effectiveness of \mname. 

\sepblock
{\bf Tiering at Host and comparison}: We evaluate \mname with three different host-level tiering solutions (enabled one at a time): Memtierd~\cite{memtierd}, TPP~\cite{tpp}, and vTMM~\cite{vtmm}.
We also compare against HugeScope~\cite{hugescope}, the current state-of-the-art for addressing page skewness in a virtualized setup. HugeScope is built on top of vTMM~\cite{hugescope} \newtext{and uses Idle page tracking to build access profile}.

Memtierd performs memory tiering using Idle Page Tracking (\ipt)~\cite{idle-page-tracking} and runs entirely in user space to perform tiering, enabling tiering even in the absence of memory pressure. TPP is part of the Linux mainline kernel and combines the LRU mechanism to detect cold pages and NUMA faults to detect hot pages. vTMM leverages Intel's Page-Modification Logging (PML)~\cite{intel-pml} to track memory accesses by guests at the host to perform memory tiering. Memtierd and TPP require no guest-side support, whereas vTMM requires guest support.

\sepblock
\textbf{Guest:} We use the KVM hypervisor~\cite{kvm} to run guest(s) with the same Linux kernel version as in the host. Each guest runs a single application instance, configured with 20\,GB of memory and 12\,vCPUs.
\newtext{For GPAC with Idle bit trackign approach,} we restrict guests to using only 4\,KB pages for \mname consolidation as discussed earlier (\S\ref{sec:impl_filter}).
\newtext{The artificat for Demeter~\cite{demeter} is supported for Icelake machine with Intel Optane. We tried running this on our Emeral Rapid server; however, it did not work as it requires a careful Linux Kernel tweaking. Since, \mname is independend of the underlying telemetry technique, in this paper we evaluation \mname using \ipt. We leave the evaluation of \mname with Intel PEBS as a future work.} 
To avoid network bottlenecks, both server and client execute in the same guest.
We refer to guests using 4\,KB base pages and 2\,MB huge pages as \baseguest and \hugeguest, respectively, in the rest of the paper. The host \textit{always} uses 2\,MB huge pages.

\begin{table}[t]
\centering
\footnotesize
\caption{Experimental platforms used in our evaluation. 
}
\label{tab:hosts}
\begin{tabular}{L{1.24cm}|L{3cm}|L{3cm}}
\hline
& \multicolumn{1}{c|}{\hostOne} & \multicolumn{1}{c}{\hostTwo} \\ \hline 
CPU & Xeon Gold 6554S & Xeon Gold 6252N \\ 
& (Emerald Rapids) & (Cascade Lake) \\ 
Memory & 2\,TB DRAM & 64\,GB DRAM \\ \hline 
& 1\,TB CXL & 512\,GB NVMM \\ \hline 
Kernel & Linux 6.12 & Linux 5.18 / 5.4 \\ \hline 
Huge Page & \texttt{always} & \texttt{always}\\ \hline 
Tiering   & Memtierd & TPP, vTMM, HugeScope \\ \hline 
\end{tabular}

\end{table}

\begin{table}[t]
\footnotesize
    \caption{Benchmarks' description with memory footprint.}
  \label{tab:benchmarks}
    \begin{tabular}{|l|C{3.5cm}|C{1.3cm}|}
      \hline
    {\bf Workload} & {\bf Description} & {\bf Guest RSS}  \\
            \hline
       Redis~\cite{redis} & In-memory key-value store & 12.5 GB  \\
      \hline
       Memcached~\cite{memcached} & In-memory key-value store & 11 GB \\
      \hline
      Hash~\cite{hash}& Hash data structure & 8.8 GB  \\
      \hline
      Ocean\_ncp~\cite{parsec3} & Ocean simulation & 5.5 GB  \\
      \hline
      Masim~\cite{masim} & Memory access simulator & 9.8 GB \\
      \hline
    \end{tabular}      
\end{table}

\subsubsection{Workloads}
We evaluate the effectiveness of \mname using a microbenchmark and a set of real-world workloads. Table~\ref{tab:benchmarks} lists the workloads along with their guest \textit{resident set size} (RSS). For Redis~\cite{redis} and Memcached~\cite{memcached}, we populate data with a key size of 1\,KB and use the Memtier~\cite{memtier_bench} to generate requests as per a Gaussian distribution. For the Hash workload, we use Hash\_bkt\_rcu~\cite{hash}, a hash table protected by a per-bucket lock for updates and RCU for lookups. We also use Ocean\_ncp, an ocean simulation with a W-cycle multigrid solver from the SPLASH2x application in PARSEC 3.0~\cite{parsec3}.
We configure the microbenchmark Masim~\cite{masim} to access only one 4\,KB page out of 512 pages in a huge page boundary.

\begin{table}[t]
\footnotesize
\centering
\caption{Consolidation time of selected hot 4 KB pages using \limit for different workloads along with the share of each operation (see Algorithm~\ref{algo:page_consol}) during consolidation.}
  \label{tab:consol_time}
    \begin{tabular}{|C{1.4cm}|C{0.6cm}|C{1.2cm}|C{1.6cm}|C{1.3cm}|}
      \hline
     {\bf Workload} & \limit &{\bf Hot 4\,KB pages} & {\bf Skewed hot huge pages} & {\bf Consol. Time (ms)} \\
      \hline
       Redis & 48 &91,959 & 3,638 & 257  \\
      \hline
       Memcached & 78 &140,977 & 2,706 & 376  \\
      \hline
      Hash & 12 &307,480 &  3,389 & 790  \\
      \hline
      Ocean\_ncp & 290 & 360,641 & 1,402 & 802  \\
      \hline
        Masim & 2 & 4,106 & 4,102 & 15  \\
      \hline
    \end{tabular}      

\vspace{1mm}

\begin{tabular}{|l|c|c|c|c|c|}
\hline
 alloc & unmap & flush & copy & set\_pte & free \\
\hline
19.9\% & 3.2\% & 24.9\% & 13.2\% & 2.9\% & 17.9\% \\
\hline
\end{tabular}

\end{table}



\sepblock
\textbf{Default \limit values:}
As discussed earlier, we use the consolidation limit -- \limit to classify a hot huge page inside the guest as \skewedhotpage or not. We profile workloads to analyze their memory access pattern and set the application-specific \limit based on the access pattern.  Table~\ref{tab:consol_time} shows the \limit, selected hot base pages, identified {\skewedhotpage}s, and the time taken to perform consolidation. 
Figure~\ref{fig:scatter_cdf} shows the knee point for different workloads. \mname does not perform consolidation for workloads with left-skewed CDF like Liblinear and SPEC 654.Roms, in which most of the huge pages are densely hot. 
{\tt Redis} is right-skewed, where the bulk of the data is concentrated towards low values, resulting in better consolidation opportunity. In contrast, {\tt SPEC 654.Roms\_s} is left-skewed, with the bulk of the data concentrated towards high values.

\subsection{Evaluation with Single Guest Instance}
\label{sec:eval_single_guest}
We start evaluation of \mname with a single guest under no near memory pressure to quantify the extent of DRAM savings using \hostOne. All consolidation processes succeed due to sufficient available contiguous memory regions. We analyze memory pressure scenarios in~\S\ref{sec:scale_memtierd} and~\S\ref{sec:eval_scale_autonuma_tpp}. Here, we use \memtierd as the representative tiering mechanism at host to migrate identified hot pages to the \nearmemory.

\begin{figure}
        \centering
        \includegraphics[width=\linewidth]{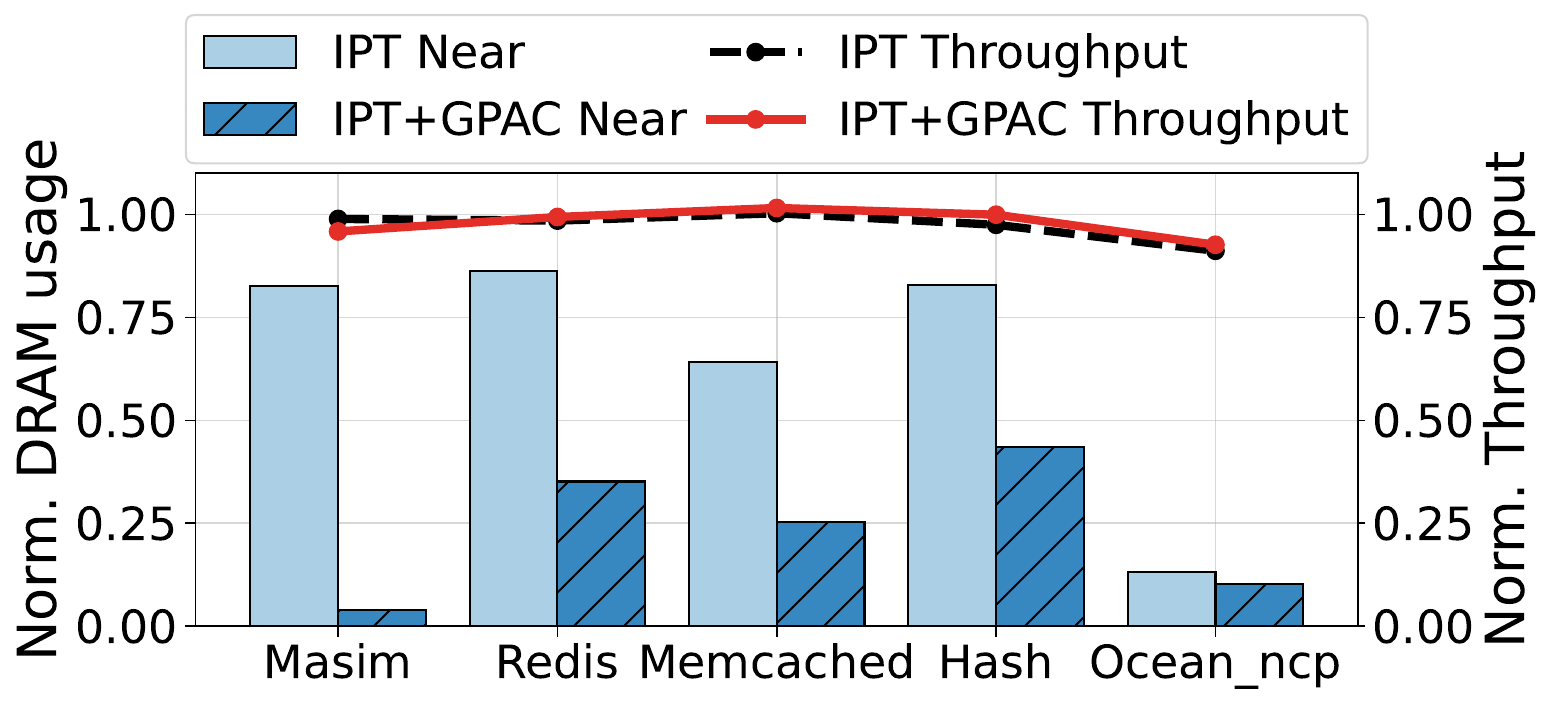}
        \caption{Reduction in DRAM usage with \mname + \memtierd compared with standalone \memtierd. \mname reduces near memory usage with minimal performance impact. Normalized to all-DRAM setting without memory tiering.}
        \label{fig:dram_reduction_and_perf}
    \end{figure}

Figure~\ref{fig:dram_reduction_and_perf} summarizes the total \nearmemory savings and performance impact across all workloads. GPAC achieves an average \nearmemory reduction of $70\%$ (excluding the microbenchmark Masim) while incurring a negligible average performance overhead of $\approx$0.86\%. \mname completely operates within the guest while the host-based tiering migrates data pages purely based on its own hot/cold classification method and policies.

We now perform a deep dive on memory tiering, and performance impact using Redis as the representative workload as it exhibits significant access skewness (see Figure~\ref{fig:scatter_cdf}).


\subsubsection{Memory Tiering impact}
\label{sec:eval_tiering_single}

\begin{figure}[t]
    \centering
    \begin{subfigure}[b]{0.49\linewidth}
        \centering
        \includegraphics[width=\linewidth]{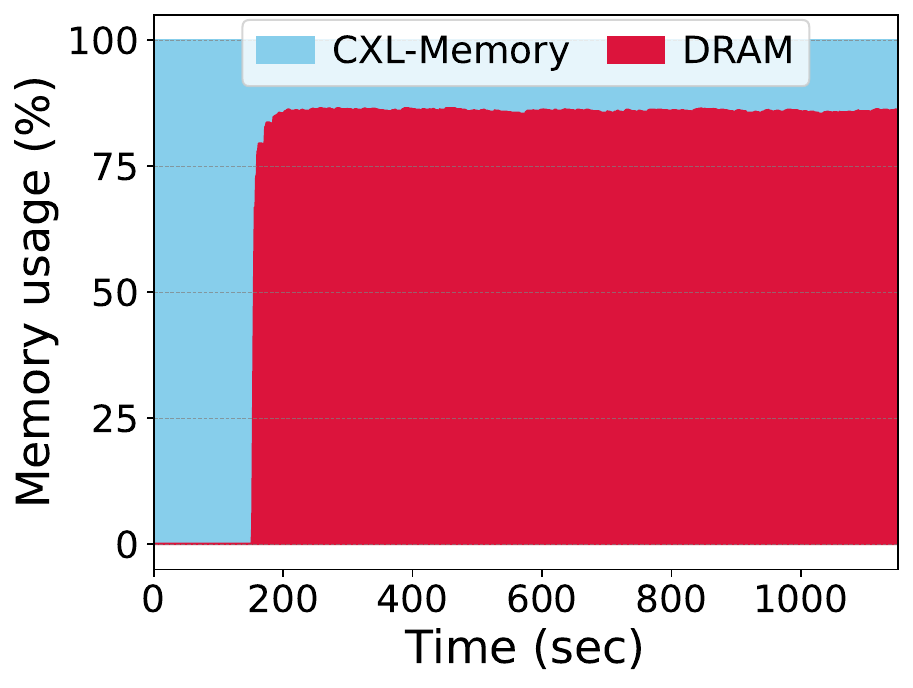}
        \caption{\memtierd}
        \label{fig:mem_baseline}
    \end{subfigure}
    \begin{subfigure}[b]{0.49\linewidth}
        \centering
        \includegraphics[width=\linewidth]{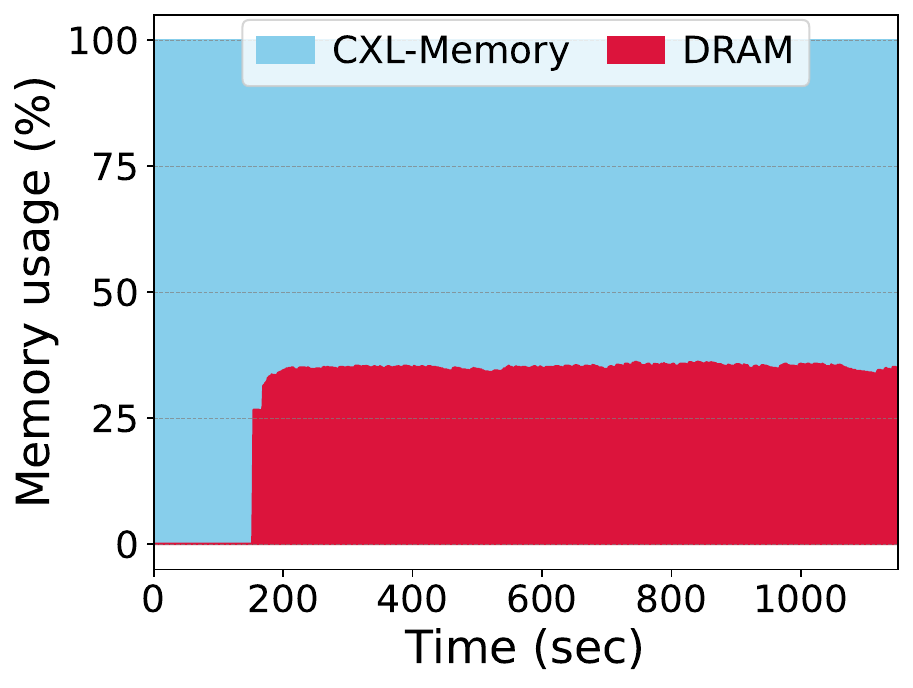}
        \caption{\memtierd + \mname}
        \label{fig:mem_gc}
    \end{subfigure}
    \caption{Memory distribution (\% of guest RSS) between DRAM and \cxl using \memtierd ($\sim$85\%) and \memtierd + \mname ($\sim$33\%) for Redis workload. }
    \label{fig:combined_mem_figures}
\end{figure}

\begin{figure}
     \centering
      \begin{subfigure}[b]{0.496\linewidth}
        \centering
        \includegraphics[width=\linewidth]{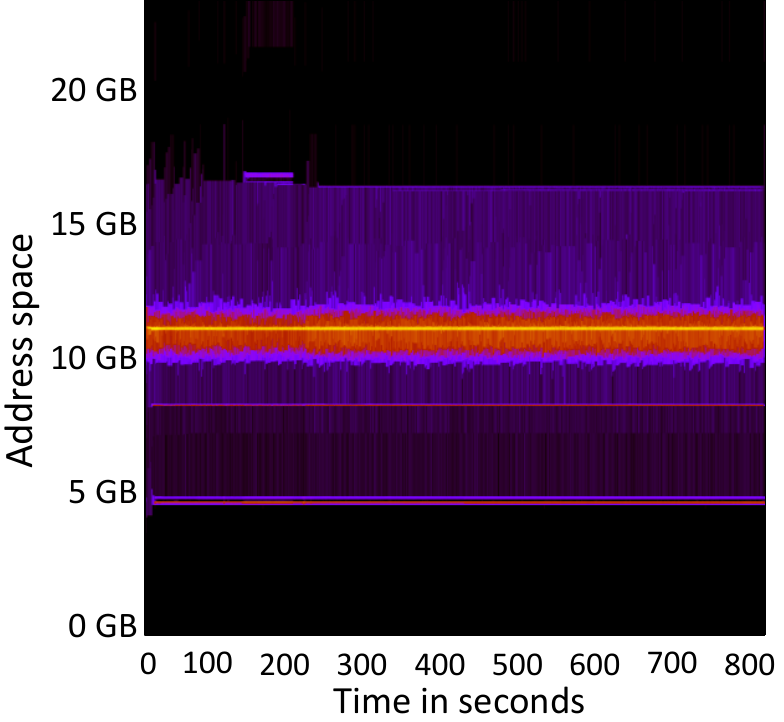}
        \caption{No consolidation}
        \label{fig:heatmap_baseline}
      \end{subfigure}
      \begin{subfigure}[b]{0.47\linewidth}
        \centering
        \includegraphics[width=\linewidth]{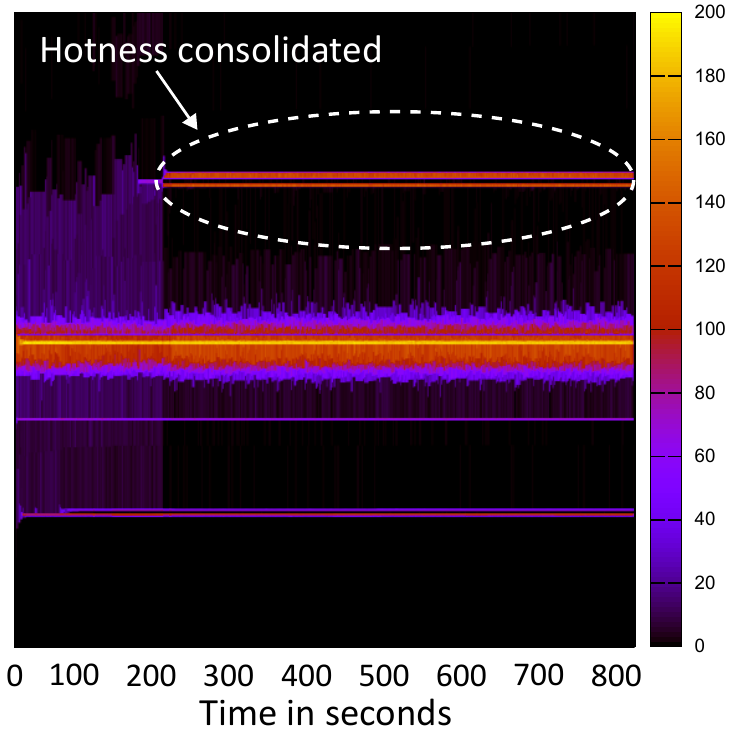}
        \caption{With consolidation}
        \label{fig:heatmap_gc}
      \end{subfigure}
        \caption{Detected hot region with Redis at host using DAMON~\cite{damon}. \mname consolidates scattered hot base pages into a few huge page regions.}
        \label{fig:heat_maps}
\end{figure}

On \hostOne, we start the Redis with \farmemory (CXL) as the {\it preferred} memory tier. When tiering begins around ~180\,seconds, \memtierd without \mname identifies ~85\% of data as hot and migrates it to \nearmemory (Figure~\ref{fig:mem_baseline}). In contrast, with \mname consolidating {\skewedhotregion}s inside the guest, \memtierd + \mname identifies only 33\% (a reduction of $\approx$ 52\%) as hot and migrates to the \nearmemory. 
Figure~\ref{fig:heat_maps} captures the impact of consolidation using DAMON~\cite{damon} to capture the detected hot regions at the host. Figure~\ref{fig:heatmap_baseline} shows the identified hot regions without \mname. The figure shows a hot region around the 11\,GB offset in the address space, which decreases in intensity (no. of accesses) as we move away from this hot region. Figure~\ref{fig:heatmap_gc} shows the detected hot regions after the consolidation with \mname. The low intensity scattered hot regions get consolidated into smaller regions. This conclusively shows the impact of \mname on the detected hot region, and eventually on the underlying memory tiering.


\subsubsection{Performance impact}
\label{sec:eval_single_performance}
The overall impact of the tiering on an application in terms of two critical metrics -- loads from \farmemory and LLC miss latency remains unaffected. In an ideal tiering solution, all the LLC misses will be served from the \nearmemory  causing a reduction in LLC miss latency (compared to a miss served from a \farmemory). We compare the number of demand data loads from \cxl and LLC miss latency (captured using \textit{perf}\footnote{{\tt unc\_cha\_tor\_inserts.ia\_miss\_drd\_cxl\_acc}, {\tt llc\_demand\_data\_read\_miss\_latency}} ) for non-tiered (DRAM-only) and tiered (\memtierd and \memtierd + \mname) setups. 

\begin{figure}[t]
    \centering
    \begin{subfigure}[b]{0.49\linewidth}
        \centering
        \includegraphics[width=\linewidth, height=.9\linewidth]{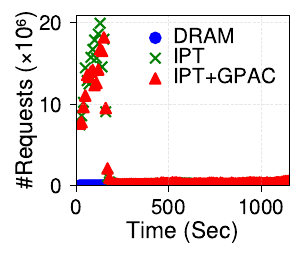}
        \caption{Load from \cxl}
        \label{fig:single_cxl_acc}
    \end{subfigure}
    \begin{subfigure}[b]{0.49\linewidth}
        \centering
        \includegraphics[width=\linewidth, height=.9\linewidth]{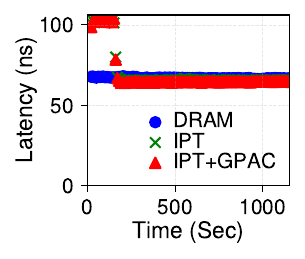}
        \caption{LLC miss latency}
        \label{fig:single_latency}
    \end{subfigure}
    \caption{Demand Loads from \farmemory (\cxl) and demand load LLC miss latency for no-tiering (DRAM), tiering with \memtierd, and tiering with \memtierd+\mname. After start of tiering around ~200 seconds, hot data is placed in the \nearmemory causing (a) a sharp drop in the Loads from the \farmemory and also a reduction in the LLC miss latency (b) for Redis.}
    \label{fig:tma_analysis_single}
\end{figure}

As shown in Figure~\ref{fig:single_cxl_acc}, the number of demand loads from \cxl before tiering begins is identical for the two tiering setups. After tiering activates, the number of \cxl accesses drops sharply and approaches DRAM-only levels. Notably, \memtierd + \mname places only 33\% of data in DRAM compared to 85\% for just \memtierd, confirming that \mname enables efficient detection of actual hot data (as the drop in \cxl loads is identical in both cases). Figure~\ref{fig:single_latency}, capturing LLC miss latency, also shows a similar pattern. We observe similar consistent behavior across other workloads.

\subsubsection{Impact of Consolidation Limit (CL)}
\label{sec:cl_impact}
The consolidation limit (\limit) balances the trade-off between \nearmemory savings and the consolidation overhead. A higher \limit may classify more huge page regions as skewed (depending on the access pattern), triggering aggressive consolidation. Conversely, a lower \limit only classifies highly skewed regions, resulting in conservative consolidation.

\begin{figure}[t]
     \centering
         \includegraphics[width=.8\linewidth]{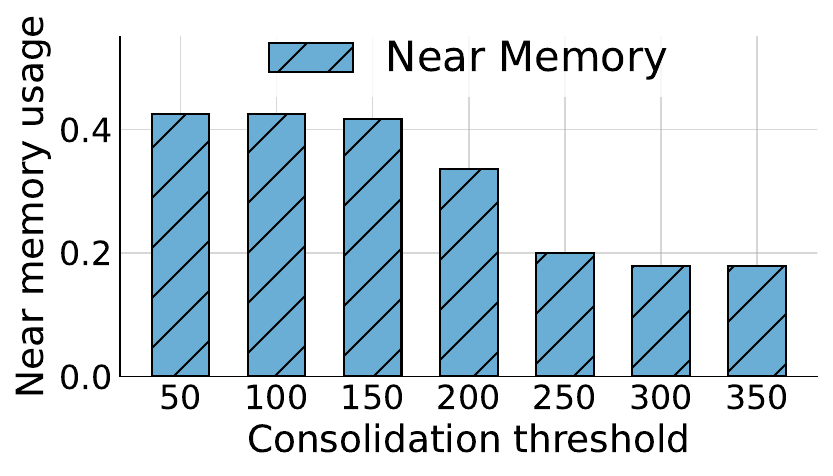}
         \caption{Impact of CL on DRAM usage for Hash workload.}        
         \label{fig:hash_threshold}
        \end{figure}

Figure~\ref{fig:hash_threshold} shows \nearmemory (DRAM) savings for the Hash workload across different consolidation limits (\limit). 
A \limit value of 50 results in a memory saving of $\approx58\%$. As CL increases beyond 150, near memory usage drops substantially, but we also observe a throughput degradation of $\approx14\%$ relative to no-consolidation performance. The performance overhead is due to aggressive consolidation process. A \limit value of more than 250 results in no additional \nearmemory savings and even no additional performance overheads, indicating only a few skewed huge page regions with more than 250 hot base pages (as confirmed by the CDF in Figure~\ref{fig:scatter_cdf}).

\subsection{At-scale evaluation using \memtierd}
\label{sec:scale_memtierd}
In this section, we measure the efficacy of \mname in a multi-tenant scenario in a \nearmemory pressure setting. In such a setting, \nearmemory used by one guest can impact the performance of other guests.
We configure multiple guests (each hosting one Redis workload with Memtier) on the same host to simulate DRAM pressure. We configure guests to prefer \nearmemory first and use \farmemory if additional memory is required. 

We use \hostOne with \memtierd tiering as the host-level tiering solution to perform memory tiering across six guests, each running the same workload. Guests are allocated a shared pool of 30\,GB DRAM and 70\,GB \cxl. For performance assessment, each guest runs Redis and Hash workloads independently with the same setting as the single VM scenario.

\subsubsection{Near memory distribution impact }
\label{sec:dram_dist_ipt}
\begin{figure}[t]
      \centering
      \includegraphics[width=\linewidth]{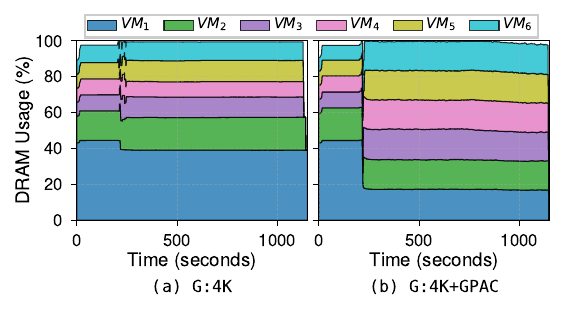}
      \caption{DRAM distribution across guests using \baseguest and \baseguest +\mname for Redis workload (\memtierd at host). 
      }
      \label{fig:stack_mem_memtierd}
\end{figure}

As shown in Figure~\ref{fig:stack_mem_memtierd}, tiering with \memtierd leads to an uneven distribution of DRAM across guests, with \textsf{$VM_1$} consuming $\approx$40\% of DRAM, while the remaining 60\% of DRAM is shared by the other 5 VMs. This uneven distribution is due to \skewedhotpage, where the majority of data pages in guests are \textit{incorrectly} identified as hot, and hence, they continue to occupy space in the already constrained \nearmemory (DRAM). Guests using either base pages (\baseguest) or huge pages (\hugeguest) suffer from the same issue.

With \mname enabled in all the guests, the number of \skewedhotpage{s} is reduced at the host, resulting in correct identification of hot and cold huge pages. Cold huge pages are subsequently demoted to \farmemory, freeing up DRAM for actual hot huge pages.
Since the DRAM capacity is sufficient to hold all the actual hot data from the guests, and as all guests run the same workload, we see an even distribution of DRAM usage.

\subsubsection{Impact on Promotion-Demotion}
\label{sec:prom_demote_ipt}
\begin{figure}[t]
    \centering
        \begin{subfigure}[b]{0.7\linewidth}
        \centering
        \includegraphics[width=\linewidth]{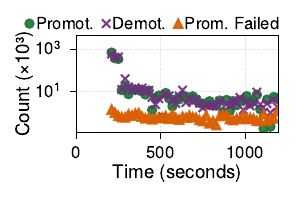}
        \caption{\baseguest}
        \label{fig:promo_demo_baseline}
        \end{subfigure}
        
        \begin{subfigure}[b]{0.7\linewidth}
        \centering
        \includegraphics[width=\linewidth]{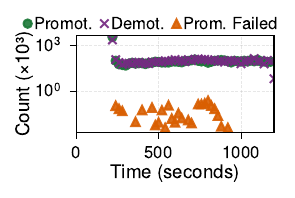}
        \caption{\baseguest + \mname}
        \label{fig:promo_demo_gc}
        \end{subfigure}

        
       \caption{Promotion/demotion of memory pages for Redis workload (Y-axis: log scale).
       }
       \label{fig:promo_demo_redis}
\end{figure}

We monitor memory pages being promoted from \cxl to DRAM and demoted from DRAM to \cxl. As shown in Figure~\ref{fig:promo_demo_baseline}, a number of promotions fail in \baseguest due to access skewness and DRAM scarcity, leading to incorrect placement of hot huge pages.
We observe similar behavior in the case of \hugeguest.
In contrast, with \baseguest + \mname (Figure~\ref{fig:promo_demo_gc}), the total number of failed promotions drops significantly, as consolidation reduces the number of \skewedhotpage{s} and frees DRAM for actual hot pages, enabling correct hot-cold placement.

\subsubsection{Impact on application performance}
\label{sec:eval_scale_ipt_perf}

\begin{figure}[t]
      \centering
      \includegraphics[width=.8\linewidth]{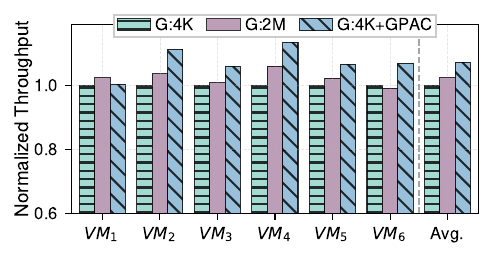}
      \caption{Performance improvement in Redis workload while using Memtierd for memory tiering at host. }
      \label{fig:redis_scale_normalized}
\end{figure}

\begin{figure*}[t]
    \centering
        \centering
        \begin{subfigure}[b]{0.31\linewidth}
        \centering
        \includegraphics[width=\linewidth]{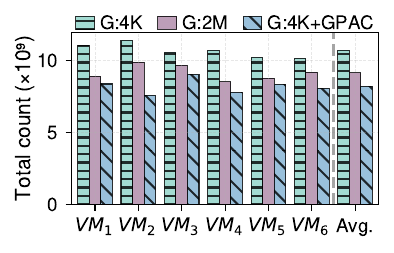}
        \caption{memory\_activity.stalls\_l3\_miss}
        \label{fig:ipt_mem_stalls}
        \end{subfigure}
        \begin{subfigure}[b]{0.33\linewidth}
        \centering
        \includegraphics[width=\linewidth]{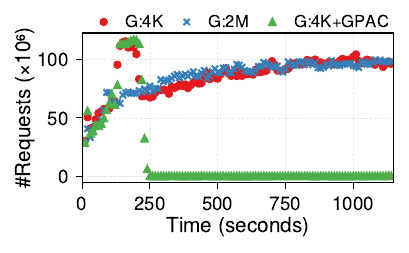}
        \caption{Demand loads served from \cxl}
        \label{fig:ipt_cxl_acc}
        \end{subfigure}
        \begin{subfigure}[b]{0.33\linewidth}
        \centering
        \includegraphics[width=\linewidth]{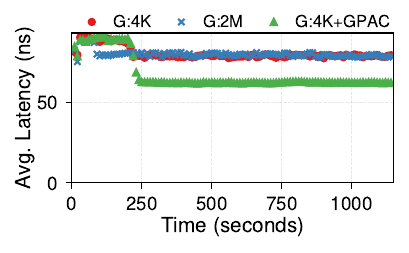}
        \caption{LLC miss latency for demand loads}
        \label{fig:ipt_latency}
        \end{subfigure}
       \caption{Impact on hardware counters such as execution stalls due to LLC miss, demand loads served from \cxl, and LLC miss latency for demand loads with \baseguest, \hugeguest, and \baseguest +\mname for Redis workload.
       }
       \label{fig:ipt_deep_dive}
\end{figure*}

Due to improved DRAM distribution, \baseguest + \mname delivers an average performance gain of $\approx$7.25\% and $\approx$4.72\% over \baseguest and \hugeguest, respectively, as shown in Figure~\ref{fig:redis_scale_normalized} for Redis.
All guests show a performance improvement of 5\% -- 12\% with  \baseguest + \mname, compared to \baseguest, with the exception of $VM_1$. For $VM_1$, we see a performance degradation of only 0.34\% even though its DRAM usage is down from $\approx$40\% to just $\approx$18\% with \baseguest, since only the actual hot huge pages from $VM_1$ are occupying DRAM.
Overall, \mname provides better system performance due to a reduction in the number of {\skewedhotpage}s at the host and a better utilization and distribution of \nearmemory.
 For the Hash workload, \baseguest + \mname improves performance by $\approx$26\% compared to \baseguest.


\sepblock
\textbf{Performance deep dive:}
We observe significant performance gain while using \baseguest + \mname, compared to \baseguest and \hugeguest. We use hardware performance counters (Table~\ref{tab:perf_counters}) to capture key statistics related to execution stalls, CXL memory accesses, and LLC miss latency.
We observed that \baseguest + \mname reduces execution stalls by 23.4\% and 10.6\% on average as shown in Figure~\ref{fig:ipt_mem_stalls}, compared to \baseguest and \hugeguest, respectively. 
Figure~\ref{fig:ipt_cxl_acc} shows a significant drop in loads served from \cxl for the entire host (by 99\%) with \baseguest + \mname, demonstrating efficient DRAM utilization. As a result, host-wide LLC miss latency for demand loads is reduced by $\approx$20.2\% with \baseguest + \mname as shown in Figure~\ref{fig:ipt_latency}.

\begin{table}[t]
\centering
\caption{Hardware performance counters used in the analysis.}
\label{tab:perf_counters}
\small
\begin{tabular}{p{3.8cm} p{3.8cm}}
\hline
\textbf{Counter} & \textbf{Description} \\
\hline
{\tt memory\_activity.\newline stalls\_l3\_miss} & Execution stalls due to outstanding LLC miss demand loads \\
{\tt unc\_cha\_tor\_inserts.\newline ia\_miss\_drd\_cxl\_acc} & Demand loads served from CXL-attached memory \\
{\tt llc\_demand\_data\_read\_\newline miss\_latency} & LLC miss latency for demand data reads \\
\hline
\end{tabular}
\end{table}

\subsubsection{Effectiveness under varying memory pressure}
\label{sec:mem_variation}
We analyze the impact of \mname under varying \nearmemory pressure using 6 guests, each running Redis. We keep the total memory fixed at 100\,GB and adjust the DRAM-to-\cxl ratio. 
At 20:80 and 30:70 DRAM-to-\cxl ratio, \memtierd+\mname outperforms standalone \memtierd by $\approx6\%--8\%$ in terms of average throughput across the 6 guests, demonstrating its effectiveness under high \nearmemory pressure. However, with 40:60 and 70:30 ratio, as DRAM availability increases, the performance gap narrows as the workloads can fit comfortably in DRAM and the advantage comes down to $\approx$4\% and $\approx$2\%, respectively.
In summary, \memtierd + \mname offers clear benefits in DRAM-constrained environments, though its advantage decreases as more \nearmemory becomes available and skewed hot pages can be accommodated without tiering.

\subsubsection{Huge pages with tiering and skewness}
\label{sec:huge_page_perf_skewness}

\begin{figure}[t]
    \centering
    \includegraphics[width=1\linewidth]{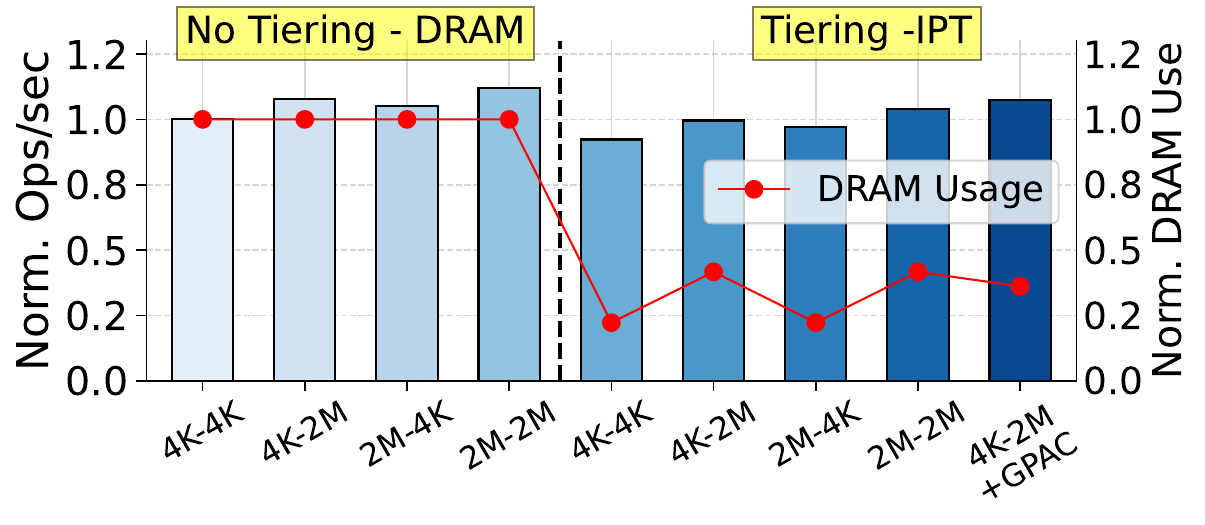}
    \caption{Performance and DRAM memory consumption of Redis for different combination of base (4K) and huge page (2M) in host and guest. Encoding: [Guest]-[Host]. Normalized to 4K-4K setting. Tiering performed based on Idle Page Tracking~\cite{idle-page-tracking}.}
    \label{fig:trend_scale_memory_norm}
\end{figure}

In a virtualized setup, in the absence of tiering, both guest and host using huge pages (\twomtwom) outperforms all other combinations by up to $\approx$20\% (Figure~\ref{fig:trend_scale_memory_norm})~\cite{page_walk_ept}.
However, as discussed earlier, using 2\,MB pages in the guest allows for only coarse-grained access tracking and does not allow for skewness detection. Therefore, GPAC uses base pages in the guest and huge pages at the host (\fourktwom). We show that, with host-level tiering enabled in the presence of \skewedhotpage{s}, \fourktwom outperforms \twomtwom in terms of performance and fast memory savings due to better utilization of fast memory

\subsection{At-scale with Various Tiering Solutions}
\label{sec:eval_scale_autonuma_tpp}

In this section, we show the performance of \mname with different state-of-the-art tiering solutions at the host: TPP~\cite{tpp}, vTMM~\cite{vtmm}, and HugeScope~\cite{hugescope}.

\subsubsection{\mname vs. vTMM and HugeScope}
\label{sec:perf_vtmm_hugescope}
\begin{figure}[t]
        \centering
        \includegraphics[width=.9\linewidth]{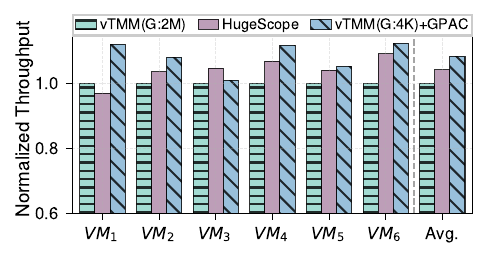}
        \caption{Performance improvement in Redis workload while using vTMM for memory tiering at host.}
        \label{fig:scale_vtmm}
\end{figure}

\begin{figure}[t]
      \centering
      \includegraphics[width=\linewidth]{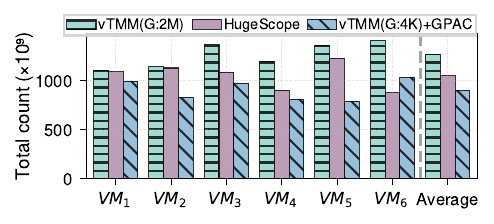}
      \caption{Stalls due to memory subsystems for Redis workload while using vTMM for memory tiering at host.}
      \label{fig:stalls_vtmm}
\end{figure}
vTMM leverages Intel PML to perform memory tiering using guest support.
HugeScope~\cite{hugescope} proposes a page splitting/coalescing policy to identify skewed hot huge pages by modifying the hypervisor. HugeScope is built on top of vTMM to perform memory tiering. We use \hostTwo with Linux kernel 5.4 in guests and host with NVMM support to compare \mname with HugeScope to maintain a similar test environment\footnote{HugeScope failed to boot on \hostOne}. To show effectiveness using vTMM and HugeScope as the memory tiering technique at host, we hosted six guests running Redis workload (each with 12.5\,GB RSS as shown in Table~\ref{tab:benchmarks}) to create pressure on DRAM. Guests are allocated a shared pool of 20\,GB DRAM and 80\,GB NVMM.

\sepblock
\textbf{Performance comparison:} We compare the performance of \baseguest + \mname with vTMM against both \hugeguest with vTMM and HugeScope. HugeScope improves performance by an average of $\approx$4.1\% over \hugeguest with vTMM. However, \baseguest + \mname with vTMM improves performance by an average of $\approx$8.5\% and $\approx$4.2\% over \hugeguest with vTMM and HugeScope, respectively, as shown in Figure~\ref{fig:scale_vtmm}. 

\sepblock
{\bf Promotion-demotion and stall analysis:}
We are unable to collect promotion-demotion stats for HugeScope and vTMM as they do not integrate the memory tiering solution with the existing NUMA interface in the Linux kernel. However, we performed analysis of execution stalls due to memory subsystems at the host level using a hardware performance counter (cycle\_activity.stalls\_mem\_any). 
We observe that HugeScope and \baseguest + \mname with vTMM reduce stalls due to memory subsystems, compared to \hugeguest with vTMM, by an average of $\approx$16.6\% and $\approx$28.4\%, respectively, as shown in Figure~\ref{fig:stalls_vtmm}.

\subsubsection{\mname vs. TPP}
\label{sec:perf_tpp}
\begin{figure}[t]
        \centering
        \includegraphics[width=.8\linewidth]{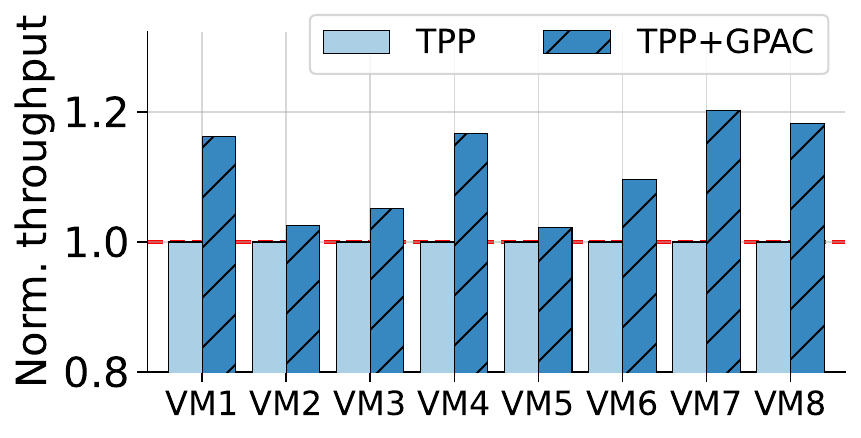}
        \caption{Performance improvement in Redis workload while using TPP for memory tiering at host.}
        \label{fig:scale_tpp}
\end{figure}

TPP~\cite{tpp} is a state-of-the-art memory tiering solution available in the Linux kernel. TPP performs memory tiering only under memory pressure by demoting cold pages to \farmemory tiers and promoting hot pages to the \nearmemory tier.
 We use the same machine (\hostTwo) to show the effectiveness of \mname using TPP as the memory tiering technique at host. We hosted eight guests running Redis workload to create pressure on DRAM. We need additional guests as TPP only tiers in extreme memory pressure scenarios.

\sepblock
\textbf{Performance comparison:} \baseguest + \mname with TPP improves the performance by an average of $\approx 11\%$ for all eight guests over \baseguest with TPP as shown in Figure~\ref{fig:scale_tpp}. The performance of VM7 and VM8 improves by 20\% and 18\%, respectively.

\begin{figure}[t]
    \centering
        \centering
        
        \begin{subfigure}[b]{0.48\linewidth}
        \centering
        \includegraphics[width=\linewidth]{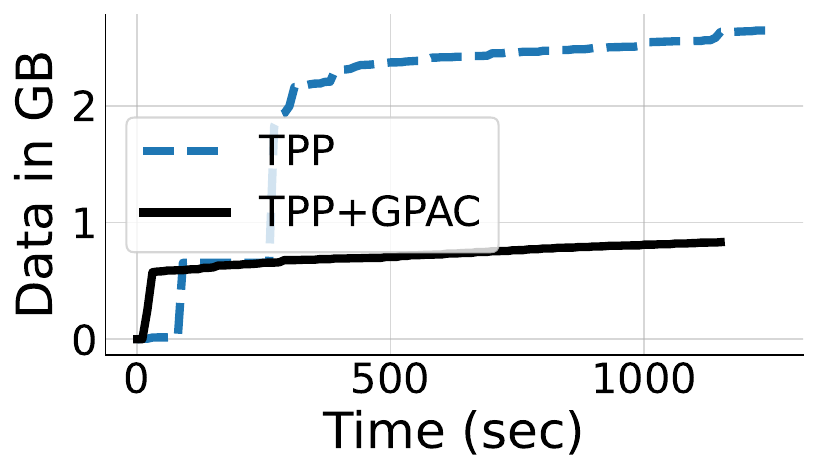}
        
        \caption{Data promoted}
        \label{fig:tpp_promote}
        \end{subfigure}
        \begin{subfigure}[b]{0.48\linewidth}
        \centering
        \includegraphics[width=\linewidth]{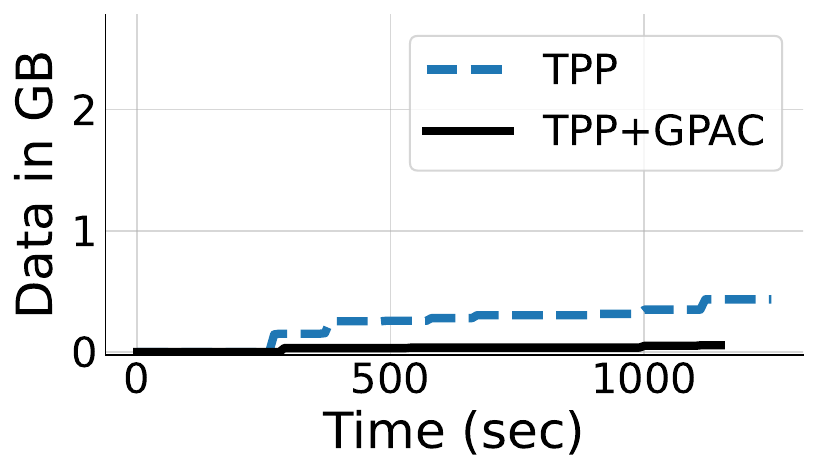}
        \caption{Data demoted}
        \label{fig:tpp_demote}
        \end{subfigure}
       \caption{Performance improvement while using TPP for memory tiering at host with data promotion and demotion.}
       \label{fig:scale_autonuma_tpp}
\end{figure}

\begin{figure}[t]
        \centering
        \includegraphics[width=.8\linewidth]{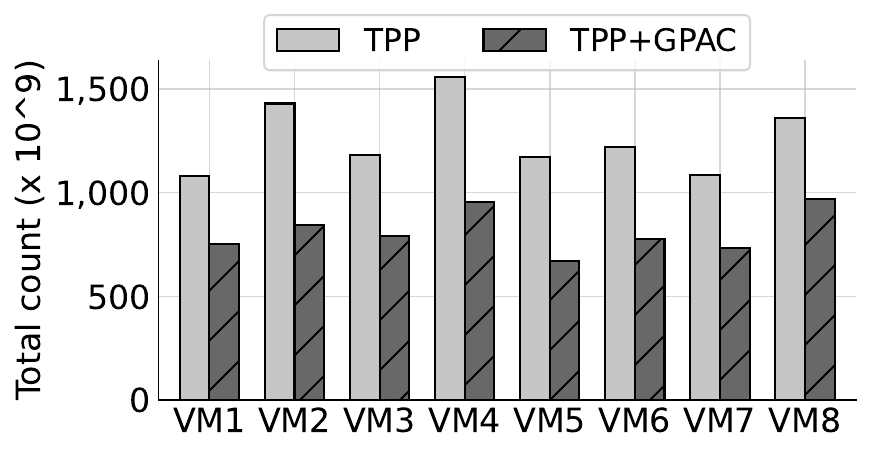}
        \caption{{Stalls due to memory subsystems for Redis workload while using TPP for memory tiering at host}}
        \label{fig:cycle_activity_stalls_mem_any_tpp}
\end{figure}

\sepblock
{\bf Promotion-demotion and stall analysis:}
Figure~\ref{fig:tpp_promote} and Figure~\ref{fig:tpp_demote} show the data promoted from NVMM to DRAM and data demoted from DRAM to NVMM, respectively, with \baseguest and \baseguest + \mname with TPP. We observe the \baseguest + \mname with TPP reduces the total data promoted and demoted by an average of $\approx64\%$ and $\approx87\%$, respectively, due to the consolidation of hot pages inside the guest to a few sets of hot huge pages at the host level. In case of TPP, we observed significant reduction in stalls by up to 42\% as shown in Figure~\ref{fig:cycle_activity_stalls_mem_any_tpp}.


\section{Conclusion}
\label{sec:conclusion}

Efficient memory tiering reduces Total Cost of Ownership by minimizing expensive \nearmemory usage. We introduced \mname, a guest physical address space consolidation mechanism that reduces the number of \skewedhotpage{s} at the host by consolidating scattered hot base pages within the guest. \mname is host-agnostic, requires no changes to host-side tiering mechanisms, the hypervisor, or hardware, and is compatible with any telemetry technique and memory tier.

\mname introduces a tunable consolidation limit (CL) that allows users to control the trade-off between \nearmemory savings and performance overhead. For a single VM, \mname reduces \nearmemory consumption by 50--70\% with minimal performance overhead. At scale, across different memory technologies, physical machines, and host-level tiering solutions such as Memtierd (\ipt), vTMM (with and without HugeScope), and TPP, \mname improves application performance by 4--11\% while significantly lowering \nearmemory usage. Notably, \mname with base pages inside the guest outperforms huge pages inside the guest under \nearmemory pressure, due to correct placement of densely hot huge pages in \nearmemory.



\newpage
\bibliographystyle{plainnat}
\bibliography{refs}

\end{document}